\documentclass[sigconf]{acmart}

\usepackage{balance}
\usepackage{graphicx}
\usepackage{amssymb}
\usepackage{amsmath}
\usepackage{xspace}
\usepackage{pdfpages}

\newcommand{\ie}{i.e., \@}

\newcommand{\cyberbunker}{CyberBunker\xspace}
\newcommand{\cyberbunkers}{CyberBunkers\xspace}

\newcommand{\cbprefix}[1]{\texttt{#1}}

\newcommand{\sref}[1]{\S~\ref{#1}}

\newcommand{\afblock}[1]{\noindent{\textbf{#1}}}

\usepackage{acronym}
\acrodef{BPH}[BPH]{\emph{Bulletproof Hoster}}
\setcopyright{rightsretained} 

\begin{document}

\title{CyberBunker 2.0 - A Domain and Traffic Perspective\\on a Bulletproof Hoster}

\author{Daniel Kopp}
\affiliation{
	\institution{DE-CIX}
}
\email{daniel.kopp@de-cix.net}

\author{Eric Strehle}
\affiliation{
	\institution{Brandenburg University of Technology}
}
\email{eric.strehle@b-tu.de}

\author{Oliver Hohlfeld}
\affiliation{
	\institution{Brandenburg University of Technology}
}
\email{oliver.hohlfeld@b-tu.de}

\fancyhead{}
\settopmatter{printacmref=true}
\copyrightyear{2021} 
\acmYear{2021} 

\acmConference[CCS '21]{Proceedings of the 2021 ACM SIGSAC Conference on Computer and Communications Security}{November 15--19, 2021}{Virtual Event, Republic of Korea}
\acmBooktitle{Proceedings of the 2021 ACM SIGSAC Conference on Computer and Communications Security (CCS '21), November 15--19, 2021, Virtual Event, Republic of Korea}\acmDOI{10.1145/3460120.3485352}
\acmISBN{978-1-4503-8454-4/21/11}




\renewcommand{\shortauthors}{}

\def\cbDomains{{1,159\xspace}}

\def\cbDomainsTlds{{52\xspace}}

\def\cbDomainsCom{{980\xspace}}

\def\cbDomainsComRelative{{85\%\xspace}}

\def\archiveDomainSnapshots{{428\xspace}}

\def\archiveDomainSnapshotsRel{{37\%\xspace}}

\def\netrayDomainSnapshots{{57\xspace}}

\def\netrayDomainSnapshotsRel{{5\%\xspace}}

\def\totalAlexaDomainsAnalyzedHR{{4.5M\xspace}}

\def\percentageSuccessfullyLargeDataset{{73\%\xspace}}

\def\totalSubdomains{{582,230\xspace}}

\def\DomainsWithPayload{{468\xspace}}

\def\identifiedCategories{{18\xspace}}

\def\blogs{{152\xspace}}

\def\parkedDomains{{112\xspace}}

\def\noCategory{{77\xspace}}

\def\Shopping{{37\xspace}}

\def\drugs{{33\xspace}}

\def\business{{31\xspace}}

\def\other{{26\xspace}}

\def\numberOfTopCategories{{6\xspace}}

\def\percentageTopCategories{{94\%\xspace}}

\begin{abstract}
In September 2019, 600 armed German cops seized the physical premise of a \ac{BPH} referred to as \cyberbunker 2.0.
The hoster resided in a decommissioned NATO bunker and advertised to host everything but child porn and anything related to terrorism while keeping servers online no matter what.
While the anatomy, economics and interconnection-level characteristics of \acp{BPH} are studied, their traffic characteristics are unknown.
In this poster, we present the first analysis of domains, web pages, and traffic captured at a major tier-1 ISP and a large IXP at the time when the \cyberbunker was in operation.
Our study sheds light on traffic characteristics of a \ac{BPH} in operation.
We show that a traditional BGP-based \ac{BPH} identification approach cannot detect the \cyberbunker, but find characteristics from a domain and traffic perspective that can add to future identification approaches.
\vspace{-.2cm}
\end{abstract}
 
\maketitle

\section{Motivation and Related Work}
Cybercrime relies on infrastructures to host their services.
One class of such infrastructures are hosting providers that promise protection, e.g., against law enforcement agencies, referred to as \ac{BPH}.
They allow cybercriminals to host any content or service while ignoring abuse messages and keeping services online.

A classical example are \emph{monolithic} \acp{BPH} which hold their own IP space.
Due to abuse generated by the hosted services, they can be detected by reputation based methods such as \textit{BGP Ranking}~\cite{bgp_rank_git} and frequent changes in the used upstream providers (e.g., due to contract terminations following abuse complaints) by analyzing BGP routing data~\cite{Konte2015}.
To avoid blacklisting, some \ac{BPH} changed their business model and evolved by abusing legitimate hosting providers \cite{goncharov2015}.
This new class of so called \emph{agile} \ac{BPH} can be detected by using whois snapshots and passive DNS data~\cite{alrwais2017}.
In May 2018, the agile \ac{BPH} \textit{Maxided} has been taken down by Dutch and Thai law enforcement agencies.
The confiscated data enabled to understand the anatomy and economics of these BPH~\cite{noroozian2019}.
While the anatomy, economics, and interconnection-level characteristics of \acp{BPH} are studied, the characteristics of the generated Internet traffic are still unknown---an aspect on which this poster aims to shed light.

The \cyberbunker 2.0 is a recent and prominent example of a monolithic \ac{BPH}; a hoster that resided in a decommissioned NATO bunker in Germany since 2013.
It advertised its service to host everything but ``child porn and anything related to terrorism'' while ``keeping servers online no matter what'', see their Stay Online Policy~\cite{CyberbunkerStayOnlinePolicy}. 
The \cyberbunker was accused to be involved in hosting illegal Internet services, see the indictment by the prosecutor~\cite{cb_indictment}.
Examples include the darknet marketplace ``Flugsvamp'', which allegedly covered around 90\% of the Swedish online drug trade~\cite{cb_indictment}.
In September 2019, 600 armed German cops seized the physical premise of the \cyberbunker~\cite{register2019}.
A post-mortem analysis~\cite{sanswhitepaper} after the take down confirms the hosting of (C\&C) servers infrastructure for multiple botnets. 
This renders the \cyberbunker as an interesting example for studying properties of \ac{BPH}.

In this poster, we aim at closing this gap by presenting the first analysis of domains, web page contents, and traffic captured at a major tier-1 ISP and an IXP at the time when the \cyberbunker was in operation.
Our study sheds light on traffic characteristics of a bullet proof hoster in operation.
We show that a BGP-based BPH identification approach would not detect the \cyberbunker and find characteristics from our traffic perspective that adds to our understanding of this type of \acp{BPH}. %
\section{Identifying the \cyberbunker}

\afblock{Identifying \cyberbunker IP space.}
We begin by using historic DNS data (see \sref{sec:dns}) to obtain IP addresses for three domains claimed to be hosted by the \cyberbunker in the indictment by the prosecutor~\cite{cb_indictment}: www.orangechemicals.com, www.acechemstore.com, and www.lifestylepharma.com (all accused for hosting shops offering narcotics and synthetic drugs).
For all domains we find A resource records pointing to IP prefix \cbprefix{B} (AS62454 - ``ZYZTM'') in the time the \cyberbunker was active.
A follow-up query in archived Spamhaus's Block List (SBL)~\cite{Spamhaus} database for the ZYZTM network yield two further prefixes \cbprefix{A} and \cbprefix{C}.
The SBL entries state the ``Hoster known to Spamhaus to be/have been involved in hosting several known professional spammers and also cybercriminal types''~\cite{Spamhaus}.
We show all identified prefixes, their ASN, and upstream provider (from BGP data 2013-2019) in Table~\ref{tab:prefixes}.
A post-mortem analysis of the \cyberbunker IP space \emph{after} take down~\cite{sanswhitepaper} confirms the usage of these prefixes.
We therefore base our analysis on the IP prefixes.
\vspace{-.5cm}
\begin{table}[!ht]
\begin{tabular}{@{}llll@{}}
\toprule
& Prefix & ASN & Upstream ASes \\ \midrule
\cbprefix{A} & 185.103.72.0/22       & 29090   &  13030        \\
\cbprefix{B} & 185.35.136.0/22       & 62454   &    9002 \& 13030  \\
\cbprefix{C} & 91.209.12.0/24       &  51088   &    61180      \\ \bottomrule
\end{tabular}
\caption{\cyberbunker IP prefixes and upstream providers.}
\vspace{-.7cm}
\label{tab:prefixes}
\end{table}

\afblock{Routing perspective.}
Prior work~\cite{Konte2015} observed \acp{BPH} to frequently change/cycle their upstream providers due to abuse.
These changes of upstream providers (``re-wiring'') are observable in inter-domain routing by analyzing BGP data~\cite{Konte2015}.
We are therefore interested if re-wiring activity could have been used to detect the \cyberbunker.
For each prefix \cbprefix{A}, \cbprefix{B}, and \cbprefix{C}, we query all routing tables that are archived every 8 hours by RIPE route collectors~\cite{RIPERIS} using BGPstream.
We use all RIBs collected by route collector \texttt{rrc00} (RIPE Amsterdam), \texttt{rrc06} (Otemachi, Japan), \texttt{rrc11} (New York, USA), \texttt{rrc12} (Frankfurt, Germany), and \texttt{rrc24} (entire LACNIC region) in the period from June 2013 till September 2019 (take down).
We find no signs for re-wiring for all three prefixes, i.e., all announced AS paths continuously feature the same upstream ASes.

\section{Domain and Web Perspective}
We first take a \emph{domain} perspective on the \cyberbunker.
This perspective enables us to understand which \emph{web sites} were hosted by the \cyberbunker.
We analyze historic DNS data and extract A resource records that point to \cyberbunker IP prefixes (\ie \texttt{www.domain.tld} $\rightarrow$ \cyberbunker IP).
The resulting list of domains enables us to later classify the content of the web sites.

\subsection{Domain Perspective}
\label{sec:dns}
\afblock{DNS data set.}
We rely on weekly DNS resolutions of all $\approx 200$\,M registered .com/net/org/... domains from DNS zone files~\cite{netrayPoster} and performing DNS resolutions during 2016 and 2019 (\cyberbunker take down).
This data contains A RR queries to \texttt{domain.tld} and \texttt{www.domain.tld}.
We extract all entries where the A RR points to an IP address included in one of the three \cyberbunker IP prefixes.

\begin{figure}[!htb]
  \begin{center}
	\includegraphics[width=1.02\columnwidth]{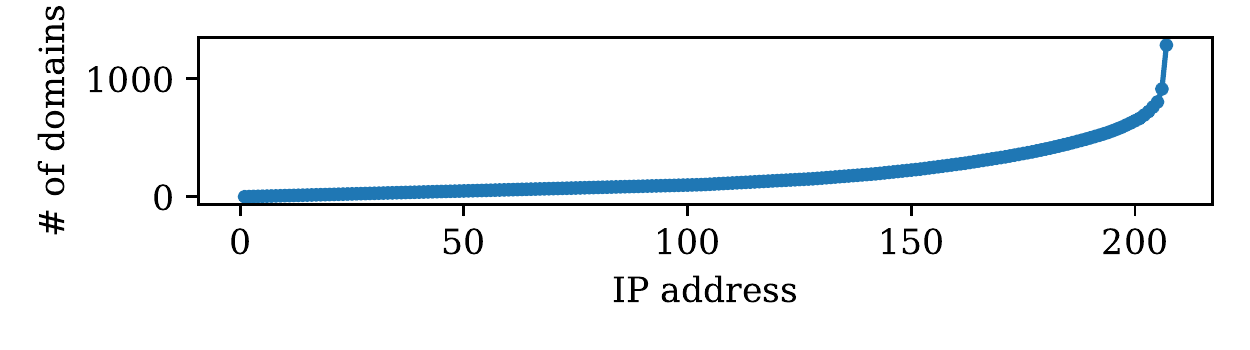}
 \end{center}
 \vspace{-.7cm}
  \caption{CDF of \#domains hosted by each used IP address.}
  \vspace{-.4cm}
  \label{fig:domain_cdf}
\end{figure}

\afblock{Domain perspective.} In total we obtain \cbDomains{} domains from \cbDomainsTlds{} different TLDs where \cbDomainsCom{} (\cbDomainsComRelative{}) domains belong to .com. These domains are located on 207 different IP addresses from \cyberbunkers IP space. Figure \ref{fig:domain_cdf} shows the Cumulative Distribution Function (CDF) for the number of domains for each used IP address. Only 18 IP addresses are used for hosting nearly 70\%\xspace of the identified domains. One single IP address is used for hosting of 372 domains. 
This analysis shows, that few IPs host the bulk of the domains and opens the question on which content is being hosted.
\vspace{-.2cm}

\subsection{Content Perspective}
Next, we aim to understand which \emph{content} was offered by web sites provided at the previously identified domains.
To answer this question, we first obtain historic snapshots (if available) for each domain and then manually classify the content of the landing page.
Since we are only interested in the offered content in the period in which the \cyberbunker was active (2013-2019), no current crawls can be used and we have to rely on archived versions.

\afblock{Data set: historic web site snapshots.}
We base this analysis on two data sets.
First, our own weekly crawls of all domains in DNS zone files since 2018, which contain the first 256K bytes of the landing page (without embedded objects such as images).
This way, we obtain HTML snapshots for \netrayDomainSnapshots{} domains (\netrayDomainSnapshotsRel).
Second, we obtain snapshots from the Web Archive (\texttt{web.archive.org}).
Using the Web Archive API, we fetch one snapshot per domain in the period of 2013 to 2019.
This way, we obtain historic snapshots for \archiveDomainSnapshots{} domains (\archiveDomainSnapshotsRel).
The intersection results to payload for \DomainsWithPayload{} domains.

\afblock{Web site classification.}
We used the OpenDNS domain classification scheme~\cite{OpenDNS} to perform a manual content classification of every domain by using 2 classifications per domain.
Some domains are not classifiable to any of the given categories. 
Therefore we extend the OpenDNS domain classification scheme with a \emph{``No Category''}-class.
We find that \percentageTopCategories{} of the web pages are represented by the top-\numberOfTopCategories{} categories, which are shown in Table~\ref{tab:classification}. 
We identified three main classes of pages: \emph{i)} domains that show no content or errors (Parked Domains; No Category) \emph{ii)} blogs with auto generated content to attract traffic (Blogs), and \emph{iii)} shops selling narcotics, drugs, or replica/fake products (Drugs; Ecommerce/Shopping).
\vspace{-.2cm}
\begin{table}[ht]
\begin{tabular}{@{}ll|ll@{}}
\toprule
Category & \#Pages & Category & \#Pages \\ \midrule
Blogs & \blogs & Drugs  & \drugs \\
Parked Domains & \parkedDomains & Business Services & \business \\
No Category & \noCategory & Other & \other \\
Ecommerce/Shopping & \Shopping & & \\ \bottomrule
\end{tabular}
\caption{Classification of web pages hosted by \cyberbunker.}
\vspace{-.9cm}
\label{tab:classification}
\end{table}

\section{Traffic Perspective}
\label{sec:traffic}
Next, we study \cyberbunker Internet traffic flows observed from a tier-1 ISP and a large IXP.
We compare and combine these perspectives to study traffic flows, protocols, and application ports to find specific features that help to characterize these types of \acp{BPH}. %

\subsection{Data Set}
We combine two vantage points: \emph{i)} a tier-1 ISP and \emph{ii)} a large IXP from which we obtain traffic flows samples of the the \cyberbunker IP space (see Table \ref{tab:prefixes}). 
The traffic flow samples do not contain any payload information and are anonymized to the extent that they cannot be attributed to individual endpoints or users.
The two dataset span a period of 6 weeks from 2019-06-17 to 2019-07-29, when the \cyberbunker was still active.
Combining both yields a volume of 18.9 TByte of transferred data in this 6-week period, which corresponds to an average traffic rate of 4.5 Mbps.

\afblock{Complementary views.}
When comparing the ISP and the IXP data set, we find two largely separate views in terms of traffic geography, but also similarities in communication patterns.
Incoming and outgoing traffic rates of the \cyberbunker network, as seen from both vantage points, correspond with a slight overweight of outgoing traffic of 68\%.
We compute a top 100 list of the source and destination networks for the IXP and ISP and find that there is almost no intersection.
This might come without surprise, but underlines the complementary view of the two datasets.
Both vantage points present similar patterns when comparing the number of source and destination networks for certain protocols. 
Most noticeable, we find a very large number of networks that are the destination of outgoing TCP connections from the \cyberbunker, compared to source networks that communicated into the \cyberbunker. 
This pattern most likely points to scanning activity outgoing from the \cyberbunker that can be seen by both vantage points, ISP and IXP.

\subsection{Application Mix}
To understand the hosted Internet applications, we study the port mix of UDP and TCP traffic flows.
First, we notice that each vantage point renders an individual port mix, when building a top list of ports sorted by flows and bytes.
Therefore, we compile the intersection of the top 100 ports from both perspectives, resulting in a clear representation of the most common and popular Internet applications related to the \cyberbunker.

As in typical Internet traffic, port 80 (HTTP) and 443 (HTTPS) for the TCP protocol and 53 (DNS) and 1194 (OpenVPN) for UDP are among the top 5 application ports used.
Extending this list to the top 10 reveals numerous application ports used for administering servers e.g. backup applications, remote desktop applications, and ports used for TOR and BitTorrent.
Most notably, we find port 22/TCP (SSH) to be one of the most prevalent application ports, even more frequent than HTTPS. 
We continue our efforts to understand the anomaly of port 22 (SSH) and find that the network of the \cyberbunker was used to scan for a vulnerability with SSH on customer edge routers towards an equivalent of more than 88,000 /24 networks. This explains our previous observed similarity in the communication pattern of outgoing TCP connections from the \cyberbunker that we observed at both the IXP and ISP.

\subsection{Traffic Characteristics}
\begin{figure}[!htb]
	 \vspace{-.4cm}
  \begin{center}
	\includegraphics[width=1.02\columnwidth]{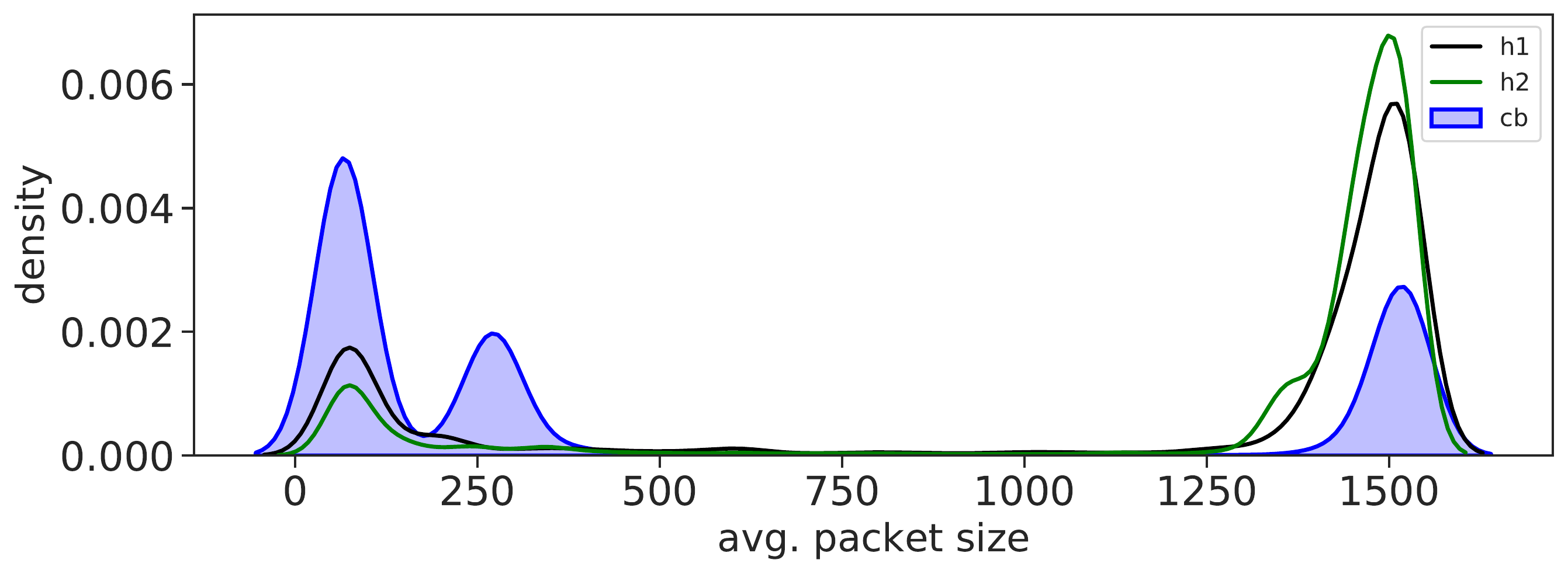}
 \end{center}
 \vspace{-.4cm}
  \caption{Packet size for HTTP/HTTPS outgoing from the \cyberbunker (cb) compared to hosting provider (h1, h2).}
  \vspace{-.4cm}
  \label{fig:cb_avg_packet}
\end{figure}

We continue to find evidence of unusual traffic patterns, especially for indications of a C\&C infrastructure that might have been hosted inside the \cyberbunker as a post-mortem analysis~\cite{sanswhitepaper} suggests.
We focus onto HTTP and HTTPS traffic as the post-mortem analysis~\cite{sanswhitepaper} indicates, from the time when the infrastructure was still active.
Indeed, we find an unusual traffic pattern that potentially points to C\&C communication when looking onto the distribution of packet sizes of HTTP and HTTPS traffic flows outgoing from the \cyberbunker.
Typically, HTTP and HTTPS communication consists of small initial packets around 64 Bytes and larger packets that transmit the content in the range of 1300 to 1522 Bytes.
In the case of \cyberbunker, we notice a large number of packets with a size between 200 and 300 Bytes, to verify that this is indeed an untypical traffic pattern we compare the distribution of packet sizes of two other major hosting providers by the help of the IXP in Figure~\ref{fig:cb_avg_packet}. This proofs, for this case, that the \cyberbunker reveals a very unusual traffic pattern.  
Further investigation of this anomaly and the targets of this unusual traffic pattern reveals around 400 /24 networks in which potential malware or bots have been contacted.

\section{Conclusion and next Steps}
In this work, we took a domain and traffic perspective on the recent case of the \cyberbunker \ac{BPH}.
We find that from well known BPH characteristics, the \cyberbunker could not be identified by upstream re-wiring in BGP. 
This difference to other \ac{BPH} make the \cyberbunker an interesting case study.
From our domain perspective, we find that web pages hosted by \cyberbunker largely differ from typical examples operational and legitimate web pages.
Additionally, the \cyberbunker shows signs of organized web hosting activity, where a few IPs host the bulk of the domains.
Our traffic perspective, shows a clear deviation of traffic patterns expected from major hosting providers and the observed patterns of C\&C-traffic corresponds with the indictment and the post-mortem analysis.
Our analysis leads to the next questions: \emph{Can the identified content of hosted websites and the discovered traffic patterns be used to identify other \ac{BPH} or is the \cyberbunker an individual case?}

\afblock{Acknowledgements.}
We thank Ingmar Poese/BENOCS for crunching traffic traces and Christopher Möller/DE-CIX for his support.
This work was partially funded by BMBF grants AIDOS 16KIS0975K and 16KIS0976.

\balance
\bibliographystyle{ACM-Reference-Format}
\bibliography{references}

\end{document}